\documentclass{sig-alternate-05-2015}

\usepackage{times}
\usepackage[absolute]{textpos}

\def\sharedaffiliation{%
\end{tabular}
\begin{tabular}{c}}

\CopyrightYear{2017}

\begin{document}

\newtheorem{challenge}{Challenge}

\newcommand{\chapquote}[3]{\begin{quotation} \textit{#1} \end{quotation} \begin{flushright} - #2 #3\end{flushright} }
\newcounter{examplecounter}
\newenvironment{example}{%
		\refstepcounter{examplecounter}%
		\textit{Example \arabic{examplecounter}}%
		 \quad 
	}{%
}

\title{UDBMS: Road to Unification for Multi-model Data Management}

\numberofauthors{1}
\author{
	\alignauthor Jiaheng Lu$^1$, ~Zhen Hua Liu$^2$, ~Pengfei Xu$^1$, ~Chao Zhang$^1$\\
	\sharedaffiliation
	\affaddr{ $^1$ University of Helsinki, Finland  and 	$^2$ Oracle, Redwood Shore, CA, USA}\\
}

\date{12 September 2016}

\begin{textblock*}{100mm}(.1\textwidth,1.5cm)
	\textit{Visionary Paper}
\end{textblock*}

\maketitle

\begin{abstract}

	A traditional database systems is organized around a single data model that determines how data can be organized, stored and manipulated. But the vision of this paper is to develop new principles and techniques to manage multiple data models against a single, integrated backend. For example, semi-structured, graph and relational models are examples of data models that may be supported by a new system. Having a single data platform for managing both well-structured data and NoSQL data is beneficial to users; this approach significantly reduces  integration, migration, development, maintenance and operational issues. The problem is challenging: the existing database principles mainly work for a single model and the research on multi-model data management  is still at an early stage. In this paper, we envision a UDBMS (Unified Database Management System) for multi-model data management in one platform. UDBMS will provide several new features such as unified data model and flexible schema,  unified query  processing, unified index structure and cross-model transaction guarantees.  We discuss our vision as well as present multiple research challenges that we need to address.

\end{abstract}


\section{Introduction}

	 As data in all forms and sizes are critical to making the best possible decisions in businesses, we see the continued growth of demands to manage and analyze massive volume of different types of data. One of the most challenging issues in the current database ecosystems is the ``\textit{Variety}'' of the data. The data may be presented in various types and formats: structured, semi-structured and unstructured. In the case of structured data, data might be structured as relational, key-value, and graph models. In the case of semi-structured data, data might be represented as XML and JSON documents.  Datasets are often produced by different sources, and hence they natively have various models and structures. 
	
	Let us consider three application scenarios to illustrate the variety of data. First, consider an application called \textit{customer-360-view} \cite{doc/IBM/customer360}, which often  requires to aggregate multiple data sources, including graph data from social networks, document data from product orders and customer information in a relation database. Second, in Oil \& Gas industry \cite{doc/ms/hems2013}, a single oil company can produce more than 1.5TB of diverse data per day \cite{journals/corr/BaazizQ14a}. Such data may be structured or semi-structured and come from heterogeneous sources, such as sensors, GPS devices, and other measuring instruments.  Third, in health-care: North York hospital needs to process 50 diverse datasets,  including structured and unstructured data from clinical, operational and financial systems, and data from social media and public health records \cite{doc/ibm/ibm2012}.  \textit{These emerging applications clearly demand the need to manage multiple-model data  in complex, modern applications, which raises two major research challenges}.

	\noindent\textbf{Data query challenge:} Users are faced with data from various sources and they need to perform global queries. However, most of the existing databases can support only one single data model, such as relation, document, graph, or key-value model, along with a specialized query language.  The restriction to a single data model limits the range of query cases that the database can handle well.  As a result,  \textit{this calls for a novel solution to support cross-model query execution and optimization}.

	\noindent\textbf{Data consistency challenge:} Updating data from multi-model stores imposes consistency challenges. For example, consider an application in social network, which consists of users and a list of friends. Assuming that users' information is stored in JSON documents, while friend relationships are stored in a graph. If we use two separated databases to manage these two types of data, the consistency guarantees can  only be provided at the individual JSON and graph databases and it may lead to an inconsistent execution. For instance, we may have scenarios where user A deletes user B in the friend graph, but  user B is still able to access the personal information of user A in JSON files. Therefore,  \textit{this calls for  a cross-model transaction principle}.
	

This paper strives to tackle the above two challenges by envisioning a system, called UDBMS (Unified DataBase Management System), that provides both cross-model query processing and cross-model consistency guarantees in one system. In the following, we first review the related works on multi-model data management (Section 1.1).  Then we show a few new challenges and our vision in UDBMS that is not covered by existing database systems (Section 2).   Finally, we categorize research challenges.

\subsection{Related work}
\label{relatedwork}

There exist two solutions to support the processing of multi-model data: (i) \textit{polyglot persistence} \cite{conf/eurosys/GogSCGCH15,journals/pvldb/ElmoreDSBCGHHKK15} and (ii) \textit{multi-model database} \cite{doc/arangodb, doc/orientdb}. \textit{Polyglot persistence}  uses numerous databases to handle different forms of data and integrate them to provide a unified interface. The history of polyglot persistence may trace back to the federation of relational engines \cite{journals/tois/HeimbignerM85}, or distributed DBMSs, which was studied in depth during the 1980s and early 1990s. Polyglot persistence approach is similar to the use of mediators in early federated database systems. The recent researches on polyglot persistence include Musketeer \cite{conf/eurosys/GogSCGCH15}, which provides an intermediate representation between applications and data processing platforms and has the merit of proposing an optimizer for the supported applications, and  BigDAWG \cite{journals/pvldb/ElmoreDSBCGHHKK15}, which enables users to run their queries over multiple vertically-integrated systems such as column stores, NewSQL engines, and array stores. In addition, RHEEM  \cite{conf/edbt/AgrawalCEKOPQ0Z16} provides a three-layer data processing and storage abstraction to achieve both platform independence and interoperability across multiple platforms for big data analytics. Overall, the existing solutions on polyglot persistence need to integrate multiple systems to provide a unified interface, which imposes further operational complexity and cost, because the integration of multiple independent databases imposes a significant engineering and operational cost. Further, in order to answer a global query, all of the databases need to remain up, which makes the fault tolerance of the application equal to the weakest database in the stack.

\begin{figure*}
	\centering
	\includegraphics[width=0.8\linewidth]{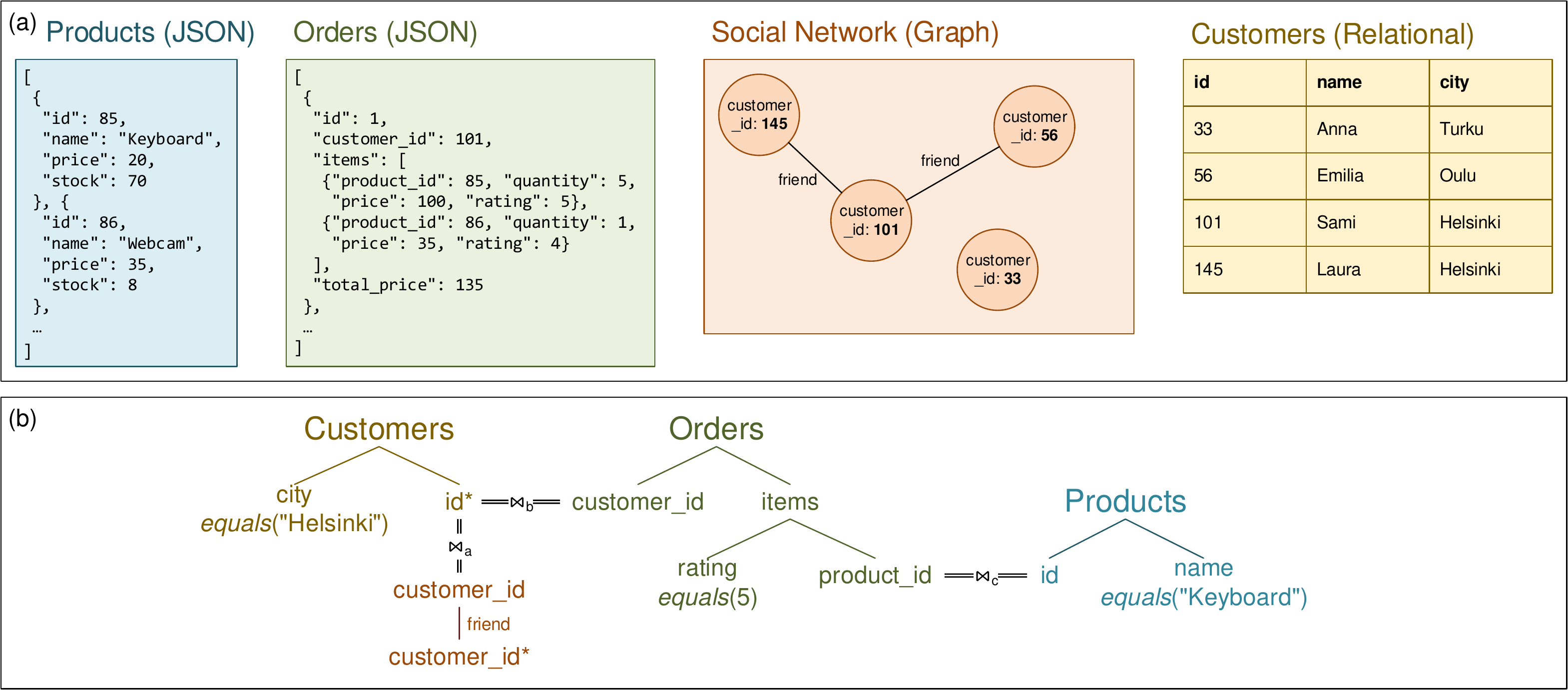}
	\caption{Example of (a) multi-model datasets and (b) query represented with a forest pattern. The elements marked with asterisk (*) are returned in query results.}
	\label{fig:join}
\end{figure*}



The second solution supports multiple data models against a single, integrated backend while meeting the growing requirements for fault tolerance, scalability, and performance. We observe a recent trend among NoSQL databases in moving away from one single model to multi-model databases. For example, OrientDB \cite{doc/orientdb} is extending graph database to support multi-model queries, and ArangoDB \cite{doc/arangodb} is moving from document model to support key-value, graph and JSON. Although the growing efforts have been put on multi-model database systems, the research on multi-model databases  is still at an early stage and there are a lot of open research questions unsolved, such as high-performance cross-model algorithms, global consistency guarantee and efficient implementation. Also, there is still lack of a global data model that can be compatible to diverse data sources. Without a (logically) unified data model,  it is hard to define global operations to  query and update different types of data in one system. In this paper, we envision a unified  system with the attempt to define a unified model and devise efficient algorithms to enable holistic query processing and transaction guarantees.

In addition, it is worthy to mention that, ``\textit{One size cannot fit all}'' \cite{conf/icde/StonebrakerC05} was proposed to argue that SQL analytics, real-time decision support, and data warehouses cannot be supported in one engine. We understand that the rationale of this argument is to encourage out-of-box thinking, that means, we cannot rely solely on RDBMS to handle a variety of new challenging requirements that do not fit the original relational paradigm. However, it is desirable to present a unified system  which hides the complexity of multiple architectures instead of having users to manage multiple systems. Therefore, in this paper, we  present our visions to significantly extend RDBMS by building a new system to query, index and update multi-model data in a unified fashion.

\section{Road to unification - Our Vision}
\label{sec:vision}

The fundamental model of a unified data management system can
be traced back to  Object-Relational DataBase Mangement System (ORDBMS), which integrates object-oriented features into a relational model.  An  ORDBMS system
 can manage  different types of data such as  relational, object, text and spatial by plugging domain specific
 data types, functions and index implementations into the DBMS kernels. For instance, PostgreSQL supports relation, spatial and XML data. Oracle continues the OR efforts to support XML, JSON and graph data.    Although, similar to our unified DBMS, the goal of an ORDBMS is to manage different types of data in one platform,  we envision a few new features and challenges in UDBMS that are not covered by the current ORDBMS architecture. In this section we lay out  these new visions and discuss  how we approach each of them respectively.
 
 


	\subsection{Data model and and flexible schema}
	
	\noindent \textbf{Unified data model.} In pure relational model, column of a table has to be a built-in scalar type. Therefore, pure RDB model is a set of  elements where each
	element is of the built-in scalar types. On the other hand, the object DB  model in  ORDBMS still has the top level set concept
	(which is commonly known as object  collection). The object collection is  the set to contain elements of
	arbitrary complex object. The object itself is yet another a set of its elements, each of which
	can be another set.  However, the current challenge in multi-model data management is that each object, such as XML, JSON and graph, models its own domain data and has specific query language (e.g. SQL for relational data, XQuery for XML and SPARQL for RDF).  Therefore, a unified multi-model DB needs to provide   a new (logically) unified data  model, which acts as a global view for different types of data. Such abstraction can hide the implementation details of data from users and facilitate the global access and query for different types of data.  Our current efforts into this direction are to unify five types of data, including {relation}, {key-value}, {JSON}, {XML} and {graph}. This goal can be achieved by two phases. At the first phase, we  envision a flexible way to represent graph, JSON, XML and key-value models as a Unified NoSQL Model (UNM) logically. At the second phase, new approaches will be investigated  to bridg the UNM and  relation models. Ultimately,  a unified model  can support all the five types of data.  This model will define global views and operations for five types of data. This unified data model will lay general foundations for accessing and manipulating multi-model data.  

\noindent \textbf{Flexible schema management.}   	Original ORDBMS assumes the perfect schema based world. Semi-structure data and unstructured data  challenges ORDBMS with schema-less design. We understand the  value of NoSQL point of schema-less DB development, but further enhances it
to argue that schema-less for write is half of the story, schema-rich for query is the other half
of the story via auto-schema derivation.  Enhancing schema discovery \cite{conf/sigmod/LiuHMLC16} for all kinds of data is
a challenge, and it is another interface that the original ORDBMS misses. There is no schema discovery indexing interface in the ORDBMS.

\smallskip

\noindent\textbf{Model evolution.}  With the increasing maturity of NoSQL databases, many applications turn to store data with JSON documents or key-value representations. But their legacy data are still stored in  the traditional RDBMS. Thus, the model change may affect the  usability of queries and applications developed on the RDBMS. Therefore, in a unified multi-model database, a research challenge is how to perform model mapping and query rewriting to automatically handle  model evolution. Note that model evolution is a more complicated than schema evolution on RDBMS, because it involves both the attribute change and the structure change.


	\subsection{Unified query processing}

 Building on a unified data model, we need to support unified query processing across multi-model data.   The original ORDBMS does not embrace a language
to process multi-model data, nor does it address the idea of doing inter-model compilation
and optimization. In contrast, to
develop a unified query to accommodate all the data, a multi-model DB needs
to support embedding model specific query language with SQL as set query language. This
query language embedding model  is illustrated  in paper \cite{conf/cidr/LiuG15}. Alternatively, there are several existing works towards providing a global  language to query multi-model data simultaneously. For instance, SQL++ \cite{journals/corr/OngPV14}  is proposed to query both JSON native stores and relational data.  ArangoDB AOL \cite{doc/arangodb} can be used to retrieve and modify both document and graph data.  We extract a core pattern, called \textit{forest pattern}, to demonstrate the core structure in a multi-model query, as illustrated below.

	\smallskip

	\begin{example} Consider an application invoving JSON documents, a relational table and a graph data in  Figure \ref{fig:join}(a). One example query is to return \textit{the friends of the customers in Helsinki who bought a keyboard and gave a five-star feedback}. This query can be used for product recommendation. Note that there are three types of joins in the query of Figure \ref{fig:join}(b): graph-relational ($\bowtie_a$), relational-JSON ($\bowtie_b$) and JSON-JSON ($\bowtie_c$) joins. The answer of this query is two pairs of customer IDs: (101,145) and (101,56). $\qed$
	\end{example}
	
	\smallskip
	
	To process the above query efficiently, as the order of joins can significantly affect the execution time, a query optimizer should evaluate available plans and select the best one. For example, how to decide the join orders among $\bowtie_a$, $\bowtie_b$ and $\bowtie_c$ in Example 1? Therefore,  one challenge is to develop
	new algorithms to select the best query plan for a multi-model query. In addition,  statistics, such as \textit{histogram} or \textit{wavelet} can be used to provide detailed information about data distribution for query optimization. The existing statistics techniques (e.g. \cite{conf/sigmod/AlwayN16}) on RDBMS are developed based on the static relation schema, but multi-model data allow the diverse and flexible schema.  Therefore,  we envision  new dynamic statistics techniques to adapt for the frequent schema changes. 
	
		\subsection{Auxiliary structure to speed up query}
	
	\noindent \textbf{Unified index structures.}  Original ORDBMS builds up domain index for each data, cross-domain query join is done by
	doing seperate domain  index probes for each domain data, and then joining the index results which is
	typically at document object ID level.  This will work if we do inter-document object join.
	However,
	In multi-model DB, such single domain index idea needs to be
	re-visited if we want to do intra-document object join. For example, 
	to support  full text search and relational scalar data search,  we need to built up 
	search indexes to incorporate IR-style inverted lists to index various data together.  But building universal search index for all
	data models requires more deep thoughts. Existing index structures focus on a single data model, e.g. \textit{B-tree} and \textit{B+-tree} are used for relational joins, \textit{XB-tree} \cite{conf/sigmod/BrunoKS02} and \textit{XR-tree} \cite{conf/icde/JiangLWO03} are developed for XML data,  and \textit{gIndex} \cite{conf/sigmod/YanYH04} and \textit{TreePi} \cite{conf/icde/ZhangHY07} are used for graph queries, our visioned system, however, executes  queries on more than one data model. Therefore, \textit{how to index multiple data models  to accelerate operations such as cross-model filtering and join?} For example, how to support $\bowtie_a$ and $\bowtie_b$ operations in Example 1 for graph-relation and JSON-relation joins efficiently?

	 In general, we can envision three types of auxiliary structures
	that we might need to build. The first is using inverted index based search index for full-text search 
	crossing multi-model DBMS \cite{conf/cidr/LiuG15}.  The second is a relational projection of various data model so that we can use
	relational schema oriented views over multi-model data. The third is building ad hoc global indexes to capture the structural feature in tree and graph data to speedup query processing.

		\smallskip

		\smallskip

	\noindent \textbf{Multi-model main memory structure.}  As the in-memory technology going forward, disk based index and data storage model is
	constantly being challenged. Building up just-in-time multi-model data structure is
	probably another interface that the original ORDBMS misses. For relational data, the traditional database vendors have developed in-memory database products such as Oracle
Database In-Memory \cite{journals/pvldb/MishraCHLLCLSKC16}, IBM DB2 BLU Acceleration \cite{journals/pvldb/RamanABCKKLLLLMMPSSSSZ13} and SQL Server In-Memory Columnstore \cite{journals/pvldb/LarsonBHHNP15}.   For other data model, such as JSON, XML, graph and key-value, we also expect in-memory solutions to speed up query  \cite{conf/sigmod/LiuHMLC16}. NoSQL database does not support join, but UDBMS shall support
	cross-domain join.  Therefore, we envision new in-memory technology for cross-model join as well. 
	
	\smallskip
	
	\subsection{Addressing CAP theorem}
	
 RDBMS supports ACID guarantee, while NoSQL  proposes BASE as the ways for scaling and workaround on CAP theorem.  We envision a per-query choice of consistency between ACID and BASE for multi-model data, which is flexible so that the user has a clear understanding and control over the performance as well as the consistency guarantees.

	 Further, to boost the performance of transaction execution, a fine-granular isolation at different levels in multi-model data can achieve the flexibility and performance benefit. For example,  objects can be isolated in the forms of subtree locks, subgraph locks, path locks and neighbor nodes locks. Further, an effective global node labeling scheme can be developed to enable the quick jump to a particular inner data node as required in the lock manager (e.g. to support \textit{getNextSibling}, \textit{getParent} operations in a tree). 
	 
	 	\smallskip
	 Sharding is a method for distributing data across multiple machines. A relational database shard is a horizontal partition of data and each shard is held on a separate database server instance to spread load. But in the scenario of multi-model data management, do we support inter-object or intra-object sharding ?
	 It is easy to support inter-object sharding. But if a graph or a tree is a big object, then we need to consider
	 about intra-object sharding. Therfore, the distributed data sharding technology needs further investigation for multi-model data.

	\subsection{Unified multi-model benchmark }
	
  A number of benchmarks have been
	proposed that can be used to evaluate big data systems (e.g. YCSB, BigBench, TPCx-BB, Bigframe). Unfortunately,
	those general-purpose big data benchmarks are not designed for the evaluation of multi-model databases.  Note that thorough
	evaluation of multi-model database systems imposes several new challenges that need to be overcome.  First, the input and output of  existing multi-model databases are quite diverse.   Since there is no standard multi-model query language available now, publicly available implementations of benchmarking data and queries for different systems should be developed, shared, unified and optimized.  Second,   unlike the relation world, NoSQL systems follow ``\textit{data first, schema
		later or never}'' paradigm. For a rigorous
	evaluation, it must be possible to control (and systematically vary) input schema for multi-model data. And the benchmark must promote productivity by enabling the creation of
	a large number of multi-model data with varied schema using little manual effort.  Finally,  multi-model databases are supposed to support the cross-model  transaction and consistency. Therefore, novel consistency metrics which describe consistency behavior for different models
	of data must be proposed in a precise way. We are currently developing  a new benchmark \cite{conf/cidr/lu17}  to provide  a rich set of examples of multi-model data and queries that can be used to improve  algorithms in multi-model database, as well as to ease the rigorous evaluation of the diverse systems.

\section{Conclusion}
\label{sec:chal}



Currently, developers  are confronted with the hard decision when they decide how to store and manage their diverse data, as their data are born out of various processes and end up in different storage platforms. To make things more sophisticated, the ``polyglot persistence'' solution suggests developers to use several platforms within a single application, by maintaining platform-specific queries at user-side, which imposes a significant engineering and operational cost for developers, as the team needs to have experts in each database technology on hand.  Thus, there is a real urgency to provide a unified system to manage different types of data. While the road to unification is full of challenges, in this paper,  we have laid down our visions to build a new system. The  challenges can be categorized into three characteristics: diversity, extensibility and flexibility.

\smallskip

\noindent\textbf{(1) Diversity}: 	The first challenge  is the ``\textit{diversity}'' of multi-model data. The existing results for query optimization and consistency model mainly work on a single model, either structured or semi-structured data. The highly diverse nature of multi-model data  makes  a unified system complicated and fascinated.

\smallskip

\noindent\textbf{(2) Extensibility}: The second challenge is to  identify the boundary of a new system? In this paper, we envision a unified system for five types of data, i.e.  {relation}, {key-value}, {JSON}, {XML} and {graph} for unification.  A further question is how to adopt the sixth type of data such as spatial data?  This calls for the future research on the extensibility of a multi-model system. 

\smallskip

\noindent\textbf{(3) Flexibility}: 	Finally, a unified multi-model system  needs to support flexible schema and model management.
Although schema evolution is a hard problem for RDBMS, in the context of multi-model databases,
there is a new way of looking at the schema evolution problem. Multi-model DBMS supports schema-less for storage, and schema-rich for query according to the automatically derived schema. So the system does not need to evolve the data. Schema
evolution becomes relatively inexpensive view definition changes instead of full data migration.

{
	\bibliographystyle{unsrt}
	\bibliography{localrefs}
}

\end{document}